\documentclass[12pt]{iopart}

\usepackage{iopams}
\usepackage[dvips]{graphicx}
\usepackage{epsf}
\usepackage{rotating}

\newcommand{\bd     }{\begin{displaymath}}
\newcommand{\ed     }{\end{displaymath}}

\newcommand{\la     }{\lambda}

\newcommand{\KA}{\alpha_A}

\newcommand{\ZA}{Z_A}
\newcommand{\ZB}{Z_{B_i}}
\newcommand{\sZB}{\{Z_{B_i}\}}
\newcommand{\ZBi}{Z_{B_i}}

\newcommand{\sZCi}{\{Z_{C_{i\cdot}}\}}

\newcommand{\nA}{A}

\newcommand{\nBi}{B_i}

\newcommand{\sB}{\{B_i\}}

\begin{document}
\title{Colouring random graphs and maximising local diversity}
\author{S.~Bounkong, J.~van Mourik, and D.~Saad}
\address{Neural Computing Research Group,
              Aston University, Birmingham B4 7ET, UK }


\begin{abstract}
We study a variation of the graph colouring problem on random graphs of finite
average connectivity. Given the number of colours, we aim to maximise the number 
of different colours at neighbouring vertices (i.e. one edge distance) of any 
vertex. Two efficient algorithms, belief propagation and Walksat are adapted to 
carry out this task. We present experimental results based on two types of 
random graphs for different system sizes and identify the critical value of the 
connectivity for the algorithms to find a perfect solution. The problem and the 
suggested algorithms have practical relevance since various applications, such 
as distributed storage, can be mapped onto this problem.
\end{abstract}


\section{Introduction}
\label{sec:intro}

The graph colouring problem~\cite{book,jort} has received a significant level
of attention. Much of this interest stems from the fact many real-world
optimisation problems can be represented as colouring problems. In the original
formulation, given $q$ colours, we aim at finding a colouring solution such
that any two connected vertices have different colours. Here, the aim is to
maximise the number of colours at one edge distance of any vertex.

One application can be found in the field of logistics, where each vertex 
represents a storage unit. The problem is then to find how to distribute the 
different types of goods such that, at each site, any type can be retrieved 
either from the given unit or from directly adjacent storage units. The problem 
that got us interested in this problem is that of distributed data storage where 
files are divided to a number of segments, which are then distributed over the 
graph representing the network. Nodes requesting a particular file collect the 
required number of file segments from neighbouring nodes to retrieve the 
original information. Distributed storage is used in many real world 
applications such as OceanStore~\cite{oceanstore}.

It should be emphasised that the main problem we are interested to solve has 
several properties that should be taken into consideration when one considers a 
colour assignment algorithm: 1) The problem is characterised by a medium number 
of different file segments. 2) An adaptive assignment of colours may be required 
as the (arbitrary) topology continuously changes due to the emergence and
disappearance of nodes. 3) The networks considered are of moderate size, 
100-1000 nodes. All this points to an efficient and fast algorithm, that can 
handle colour assignment in large systems.

Although this problem has not yet been shown to be NP-complete, it seems 
nonetheless intractable for a large system size. Since no research has been 
carried out on this specific problem, no dedicated tools exist either\footnote{
It should be mentioned that different systems of similar topology have been 
investigated in~\cite{MB}.}. However, as we report in this paper, existing
optimisation algorithms can be adapted quite easily to solve this and similar 
problems. In particular, we investigate two well established techniques: belief 
propagation (BP) and a variant Walksat (WSAT) for this purpose.

In this paper, we show how message passing techniques (BP) can be used to solve 
this particular hard computational problem, and use the results to identify the 
transition point in terms of the connectivity above which the algorithms are 
able to find a perfect colouring for the graphs. For a given number of colours 
$q$, we identify the critical connectivity $\lambda^q_c$ above which graphs are 
typically colourable by the algorithms, as well as the average minimum measure 
for the unsatisfaction $E^q(\lambda)$ as a function of the connectivity 
$\lambda$. In a general setup, the measure of unsatisfaction is 
$E^q(\lambda)=\sum_{i=1}^nE^q_i(\lambda)$ where for each vertex $i$ with local 
connectivity $\lambda_i$
\begin{equation}
E^q_i(\lambda)=\min(q,\lambda_i+1)-q_i
\label{eq:unsat}
\end{equation}
is the difference of the maximal number of available colours (from itself and
its nearest neighbours, i.e. $\min(q,\lambda_i+1)$), and the number of actually
available colours at that node $q_i$. In this paper, we only consider graphs 
with local connectivities $\lambda_i\geq q-1$, such that $E^q_i(\lambda)=q-q_i$ 
just counts the number of missing colours. One should note that, contrary to the 
original graph colouring problem, the problem of finding a colouring for our 
problem actually becomes easier with increasing connectivity.

The main goal of this paper is to introduce the problem, and to investigate
the behaviour of two algorithms on realistic system sizes. The analysis of the
model in the sense of a phase diagram for infinite system size, is a separate
issue that is currently being investigated.

\section{Belief propagation}
\label{sec:bp}
Belief propagation, a non-local algorithm also called the sum-product algorithm,
relies on iterative message passing and provides near optimal performance at low
computational cost in a wide range of applications~\cite{pearl,mackay}. Message
passing techniques rely on conditional probabilistic messages passed from the
immediate neighbourhood to find the most probable assignment of states to
variables given the constraints. In our problem, the constraints correspond to
the fact that at each vertex, one should be able to retrieve $\min(q,\lambda_i
+1)$ colours from the vertex itself and its first order neighbours.

\begin{figure}
\centering
\begin{picture}(400,180)
\put(10,150){$(a)$}
\put(170,150){$(b)$}
\includegraphics[width=0.4\linewidth]{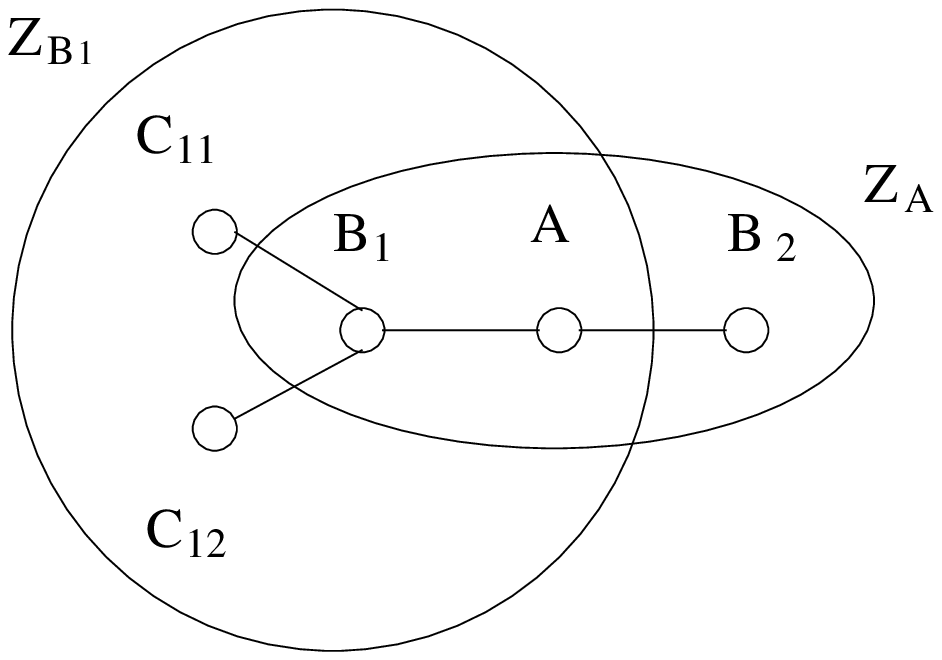}
\includegraphics[width=0.4\linewidth]{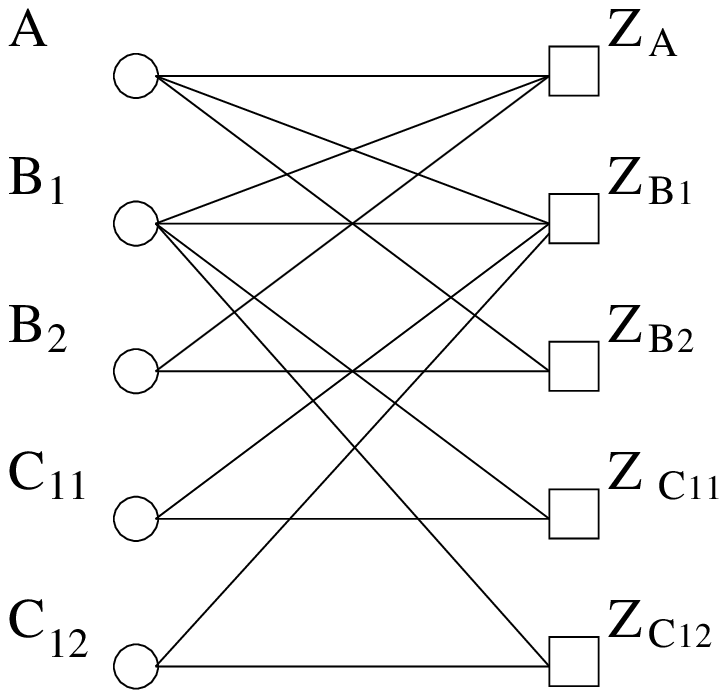}
\end{picture}
\caption{(a) A graph representing the successive local
constraints. (b) A bi-partite graph representation of the
problem.} \label{fig:graph}
\end{figure}

These constraints can be represented by clusters of vertices on a graph as in
Fig.~\ref{fig:graph}a, where `A', `$\mathrm{B}_1$', `$\mathrm{B}_2$',
`$\mathrm{C}_{11}$' and `$\mathrm{C}_{12}$' correspond to the vertices, while
`$Z_A$', `$\mathrm{Z}_{\mathrm{B}_1}$', `$\mathrm{Z}_{\mathrm{B}_2}$',
`$\mathrm{Z}_{\mathrm{C}_{11}}$' and `$\mathrm{Z}_{\mathrm{C}_{12}}$' are check
variables corresponding to the constraints. The checks are variables related to 
the unsatisfaction of a given colour assignments for a node `A' and its direct 
neighbours
\begin{equation}
P(\ZA|A,\sB) = e^{-\beta(q-q_A)}\,,
\end{equation}
where $q_A$ is the number of available colours to $A$ and $\beta$ is fixed to 
20.

The graph can be transformed into a bipartite graph, which separates the
vertices from the checks. The following update rules can be easily obtained by
naively adapting the original belief propagation rules~\cite{pearl,mackay}.
Thus, the messages from a check to a vertex are given by
\begin{eqnarray}
P(\ZA|\nA)
  & =      & \sum_{\sB}P(\ZA|A,\sB)P(\sB|\sZB,\sZCi)~,\nonumber\\
  & \simeq & \sum_{\sB}P(\ZA|A,\sB)\prod_iP(\nBi|\ZBi,\sZCi)~,
\label{PZA}
\end{eqnarray}
while the message from a vertex to a check is given by
\begin{equation}
P(\nA|\sZB) = \KA\prod_{\sZB} P(\ZB|\nA)~,
\end{equation}
Finally, the pseudo-posterior is given by
\begin{equation}
P(\nA|\ZA,\sZB) = \alpha P(\ZA|\nA)\prod_{\sZB} P(\ZB|\nA)\,,
\label{PA}
\end{equation}
where $\KA$ and $\alpha$ are normalisation coefficients. Note that the 
factorisation in (\ref{PZA}) is a relatively crude approximation even in the 
large system limit, as the $\sB$ are correlated. To deal with this properly, a 
more advanced analysis using a cluster expansion~\cite{cvm} is currently been
undertaken. Nevertheless, as we will see these approximations work remarkably
well.

If convergence of the BP algorithm is reached, the colours of vertices whose
(pseudo) posterior is greater than a pre-defined threshold set at $0.9$ in our
experiments can be fixed. If no such high posterior exists then the vertex with
the highest posterior value has its colour fixed. Then, the update rules are
re-iterated and the decimation process repeated until a global colouring is
extracted.

A major drawback of BP algorithm is that convergence is not guaranteed.
Since all the probabilities are initialised randomly, different parts of the 
graph may converge to different local solutions. Then, parts with incompatible 
colourings will continue to compete with each other, resulting in non 
convergence.

Time averaging~\cite{jff} is a way of getting around the problem by carrying out
the decimation and colour fixing process according to the average posterior
(over time i.e. a number of iterations) instead of instantaneous posterior. In 
the case of non-convergence, this method will decimate the vertex with the 
strongest {\em average} colouring probability over all competing solutions and 
thus reduce the fluctuations due to the competition. After several trials, a 
time window of 30 iterations was chosen for all numerical data presented here. 
This turned out to give the best balance between the quality of the obtained 
colourings and the computational cost.

We have opted for BP combined with time averaging, as for the ultimate task of
distributed storage we have in mind a scenario in which nodes may suddenly
switch off and turn back on again (as is often the case in peer to peer
networks). Then time averaging may have a significant benefit over other
algorithms, being able to take the average probability that a node is able to
provide a certain segment (or not), into account. This aspect however has not
been included in the current paper.

\section{Walksat}
\label{sec:wsat}
Walksat is a local search algorithm, originally designed to solve the problem 
of finding variable assignments that satisfy as many clauses as possible of a 
given conjunctive normal form~\cite{selman}. Although Walksat local search may 
seem to be suboptimal at first sight, studies have shown it to be a powerful
tool~\cite{aurell}. Many variants of the original algorithm 
exist~\cite{selman,mcallester,patterson}. In this study, we have adapted the 
variant referred to as SKC, for solving this specific colouring problem.

The original Walksat heuristic (SKC) uses the notion of the \emph{breakcount} 
of a variable, which is the number of clauses that are currently satisfied, but 
would become unsatisfied if the variable assignment were to be changed. The SKC 
variable selection is as follows:
\begin{enumerate}
\item If there are variables with breakcount equal to 0, randomly select one
  such variable.
\item Otherwise
  \begin{itemize}
  \item with probability $p$  randomly select a variable.
  \item with probability $1-p$ randomly select a variable with minimal
    breakcount.
  \end{itemize}
\item Flip the selected variable.
\item Repeat until all clauses are satisfied or until the max-iterations
    is reached.
\end{enumerate}

In our problem, the \emph{breakcount} of a variable is given by the number of
vertices for which the change of assignment would decrease $q_i$. Henceforth,
the \emph{breakcount} is now dependent on the replacement colour. In the first 
step (i) of the SKC procedure, the selected replacement colour is the one which
leads to a \emph{breakcount} equal to 0 (if more than one, choose one randomly). 
In the second step (ii), a replacement colour is selected at random. In our
first few attempts, this adaptation of the Walksat algorithm showed mixed 
results, which were up to 50\% worse than those obtained with the BP algorithm.

Therefore, we adapted another local search algorithm~\cite{cohen} related to
Walksat. This other algorithm is also iterative and based on a mixture of
gradient and ``{\em noisy}'' descent. At each iteration, one of these two
descents is chosen at random, with some probability. Similarly to the Walksat 
algorithm, this step is repeated until all checks are satisfied, or until the 
maximal number of iterations is reached.

The gradient descent is operated by the algorithm named GSAT~\cite{mitchell},
which during an iteration changes the assignment of the variable that leads to
the greatest decrease in total number of unsatisfied clauses. In our problem,
the change will be made such that it leads to the greatest decrease in
unsatisfaction as defined in (\ref{eq:unsat}). The ``{\em noisy}'' move of the
original algorithm~\cite{cohen} is replaced here by the Walksat SKC heuristic.
Therefore, the resulting algorithm is a mixture between SKC and GSAT, which is
parameterised by a probability $p_m$ that set to 0.5. Our experiments shows that
this mixture between SKC and GSAT performs significantly better than SKC alone
with no increase of the computational cost.

If not all checks are satisfied at the maximal number of iterations, the GSAT
algorithm is iterated until a local minimal is reached. The choice of maximal
number of iterations is discussed in the next section. The latter algorithm,
referred to as Walksat, shows that results are qualitatively similar to
the ones obtained using the BP algorithm both in terms of the measure of 
unsatisfaction $E^q(\la)$ and of the number of perfect colouring found.

\section{Simulations}
\label{sec:sim}
The experiments are carried out for $q=4$ colours and based on two system sizes
($n$): graphs of 100 and 1000 vertices. The studied graphs have an average
connectivity $\lambda$ and are of two types, referred to as (cut-) Poissonian
and linear graphs:
\begin{itemize}
\item
The vertices of the (cut-) Poissonian graphs with average local connectivity
$\la$  have local connectivities $\la_i$ given by
\begin{equation}
\lambda_i=\lambda_{\min}+z_{\lambda-\lambda_{\min}}=q-1+z_{\lambda-q+1}
\end{equation}
where $z_{\lambda-q+1}$ is randomly drawn from a Poisson distribution with
parameter $\lambda-q+1$.
\item
The vertices of so-called linear graphs with average local connectivity
$\la$, have local connectivities $\lambda_i$ given by:
\begin{equation}
\lambda_i=\lfloor\lambda\rfloor+z_{\lambda-\lfloor\lambda\rfloor}
\end{equation}
where  $\lfloor\lambda\rfloor$ is the largest integer smaller or equal to
$\lambda$, and where $z_{\lambda-\lfloor\lambda\rfloor}$ is $1$ with probability
$\lambda-\lfloor\lambda\rfloor$ and $0$ otherwise.
\end{itemize}
We study the most interesting range of average connectivities from $\lambda=3$
to $\lambda=5$ with a step of 0.1. For each $\lambda$, 1000 graphs of each type
were randomly generated and then coloured by both the BP and Walksat algorithms.

\subsection{Graph characteristics}
\label{sec:graph}
The graphs and the constraints are born from the original problem we have set to
solve, namely distributed storage. Here, we shall point out two observations
that may help in getting insight into the characteristics of the problem and
solutions found by the algorithms.

The number of checks is always equal to the number of vertices. Indeed, each 
vertex is associated with a check, that connects it to all vertices at one edge 
distance. This check is obeyed when the vertex can retrieve all possible colours 
from vertices at one edge distance.

The second observation is related to the fact that edges are not directed: if 
the vertex `B' is connected to the check of `A', then the vertex `A' is also
connected to the check of `B'. Hence, there are always $\frac{n\lambda}{2}$ 
short loops, which correspond to the number of edges, in the belief network 
even in the large system limit. When the connectivity value $\lambda$
increases, the number of loops increases as well, but it also becomes easier
to get a lower value for the average unsatisfaction. Therefore, it is not clear
whether the influence of the presence of loops on the performance of the
(current) BP algorithm will increase or decrease with $\lambda$.

\subsection{Walksat performance}
In the Walksat algorithm, the maximal number of iterations $nbit$ is an
important parameter. A greater value increases performance, but also
computational cost. Unfortunately, the relation between performance and cost is
not linear and it is therefore difficult to estimate the optimal number of
iterations. In order to understand this relation, we carried out several
simulations with different values of $nbit$ for the two systems sizes and all
connectivity values.

\begin{figure}
\centering
\begin{picture}(400,180)
\put(10,160){$(a)$}
\put(220,160){$(b)$}
\includegraphics[width=0.43\linewidth]{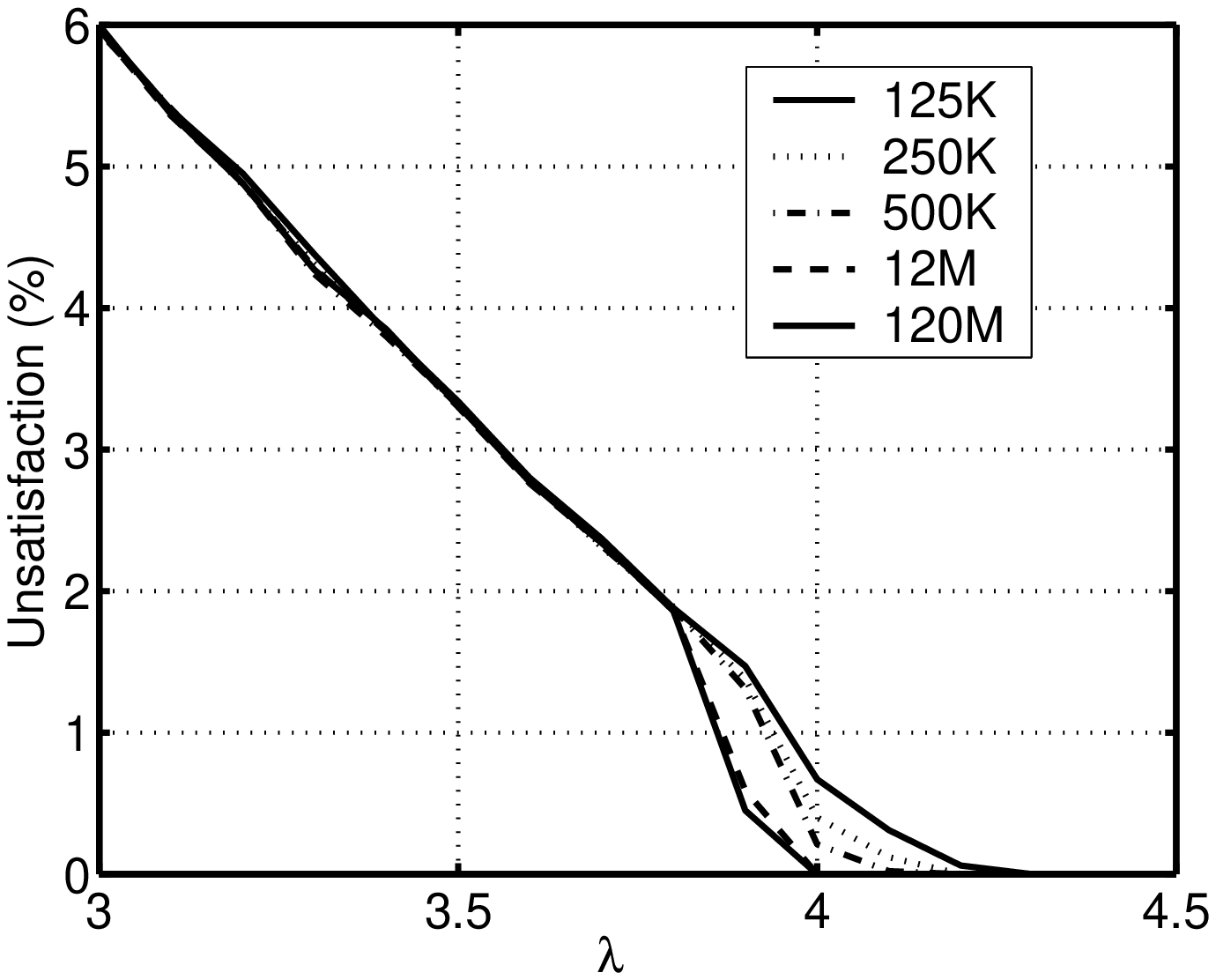}
\includegraphics[width=0.45\linewidth]{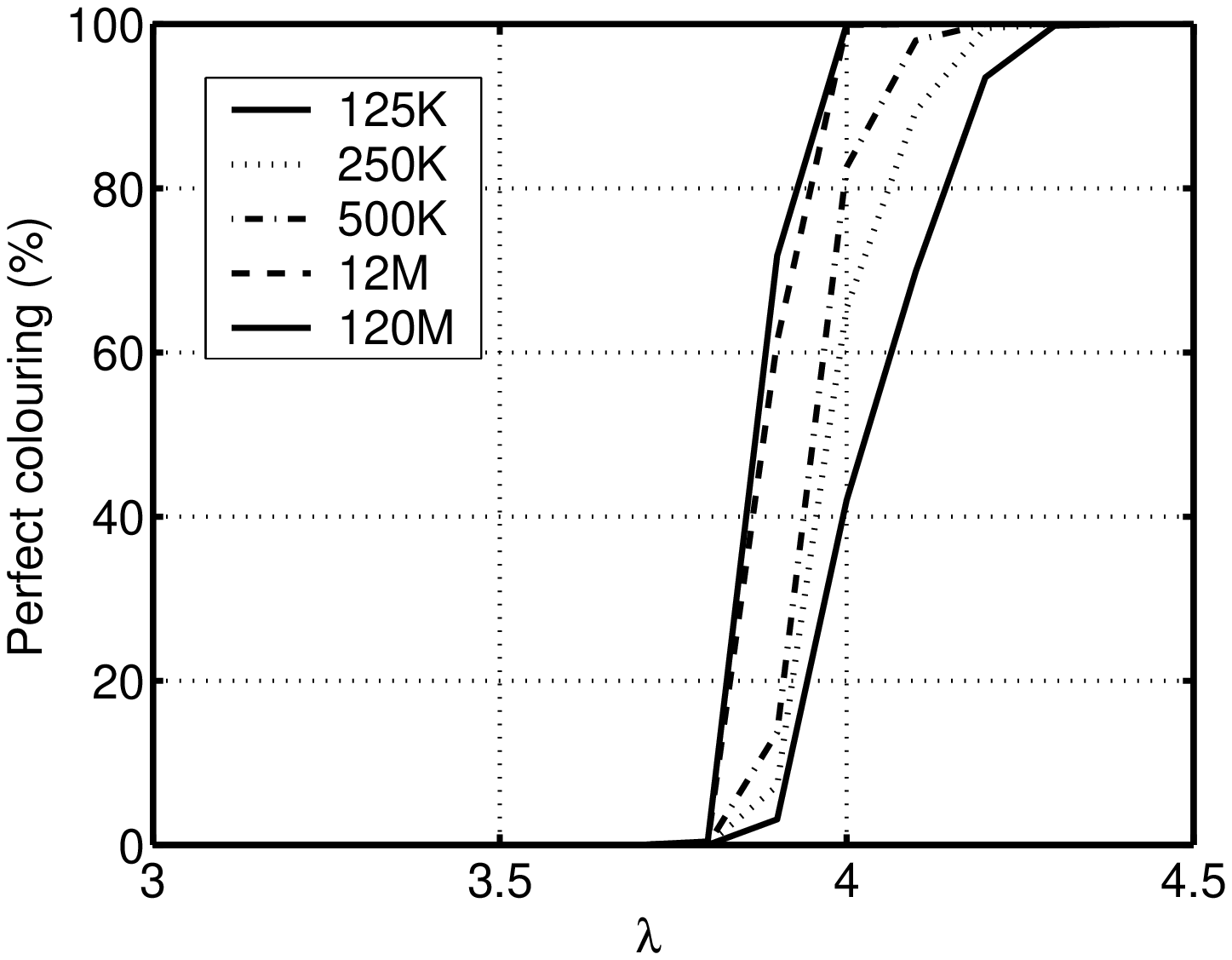}
\end{picture}
\caption{Walksat performance on linear graphs (n=100) for various $nbit$ (from
125K to 120M iterations) and connectivity values $\lambda$. (a)~The
unsatisfaction measure. (b)~Percentage of perfect colouring solutions.}
\label{wsat}
\end{figure}

\begin{figure}
\centering
\begin{picture}(400,180)
\put(10,160){$(a)$}
\put(220,160){$(b)$}
\includegraphics[width=0.43\linewidth]{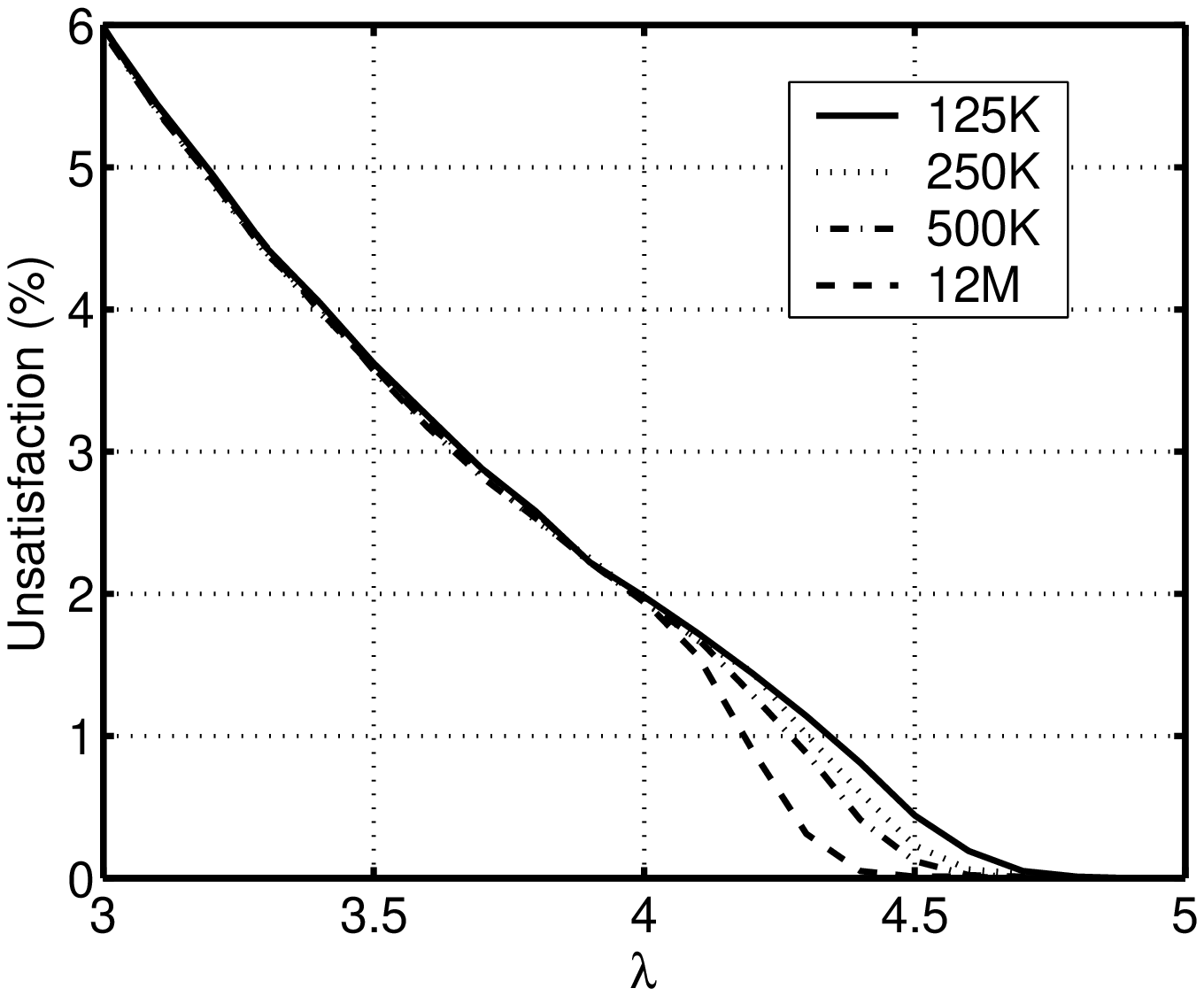}
\includegraphics[width=0.45\linewidth]{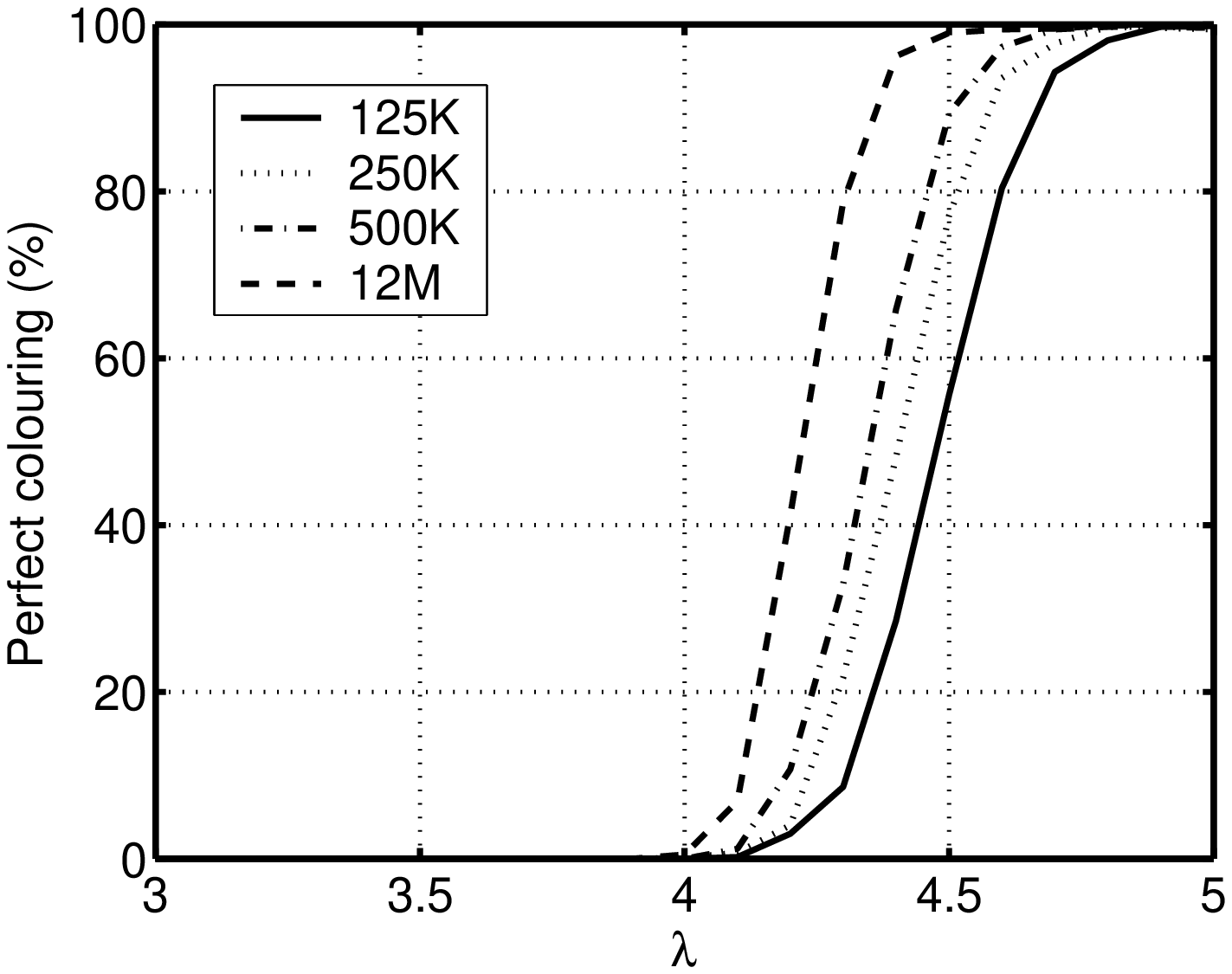}
\end{picture}
\caption{Walksat performance on Poissonian graphs (n=100) for various $nbit$
(from 125K to 12M iterations) and connectivity values $\lambda$. (b)~Percentage
of perfect colouring solutions.}
\label{wsat_pois}
\end{figure}

Figure~\ref{wsat} and~\ref{wsat_pois} show the results obtained for a system
size of 100 vertices and a range of limits on the number of iterations. One 
notices that improvements in terms of unsatisfaction and perfect colouring is
negligible for $\lambda\le3.8$ for linear graphs and for $\lambda\le4.1$ for 
Poissonian graphs. In these regions, no perfect solutions are actually found. 
Hence, the Walksat algorithm only stops when the maximum number of iterations 
is reached and returns the unsatisfaction of the nearest local minima. If the 
lowest unsatisfaction value over all examined colour assignments were returned 
instead, then one could expected improved results for larger values of $nbit$.

For $\lambda\gtrsim3.8$ and $\lambda\gtrsim4.1$ for respectively linear and
Poissonian graphs, when some perfect colouring solutions exist and are found,
increasing $nbit$ does actually decreased the unsatisfaction value. However, a
larger number of vertices will also require an exponentionally larger number
of iterations to achieve the same performance. Ultimately, if $nbit$ were
infinite, Walksat would find perfect colouring solutions.

Hence, to compare performance of the Walksat and BP algorithms, we take the
results achieved by Walksat for roughly the same computational time as the one
used by the BP algorithm. This means $nbit=500K$ and $nbit=12M$ iterations for
systems sizes of 100 and 1000 vertices, respectively. We also amend the Walksat
algorithm described in Sec.~\ref{sec:wsat} such that the unsatisfaction
returned will be the lowest unsatisfaction value over all examined colour
assignments and not the one corresponding to the nearest local minima.

\subsection{Comparison of the performance of the BP and Walksat algorithms}

\begin{figure}
\centering
\begin{picture}(400,180)
\put(10,160){$(a)$}
\put(220,160){$(b)$}
\includegraphics[width=0.43\linewidth]{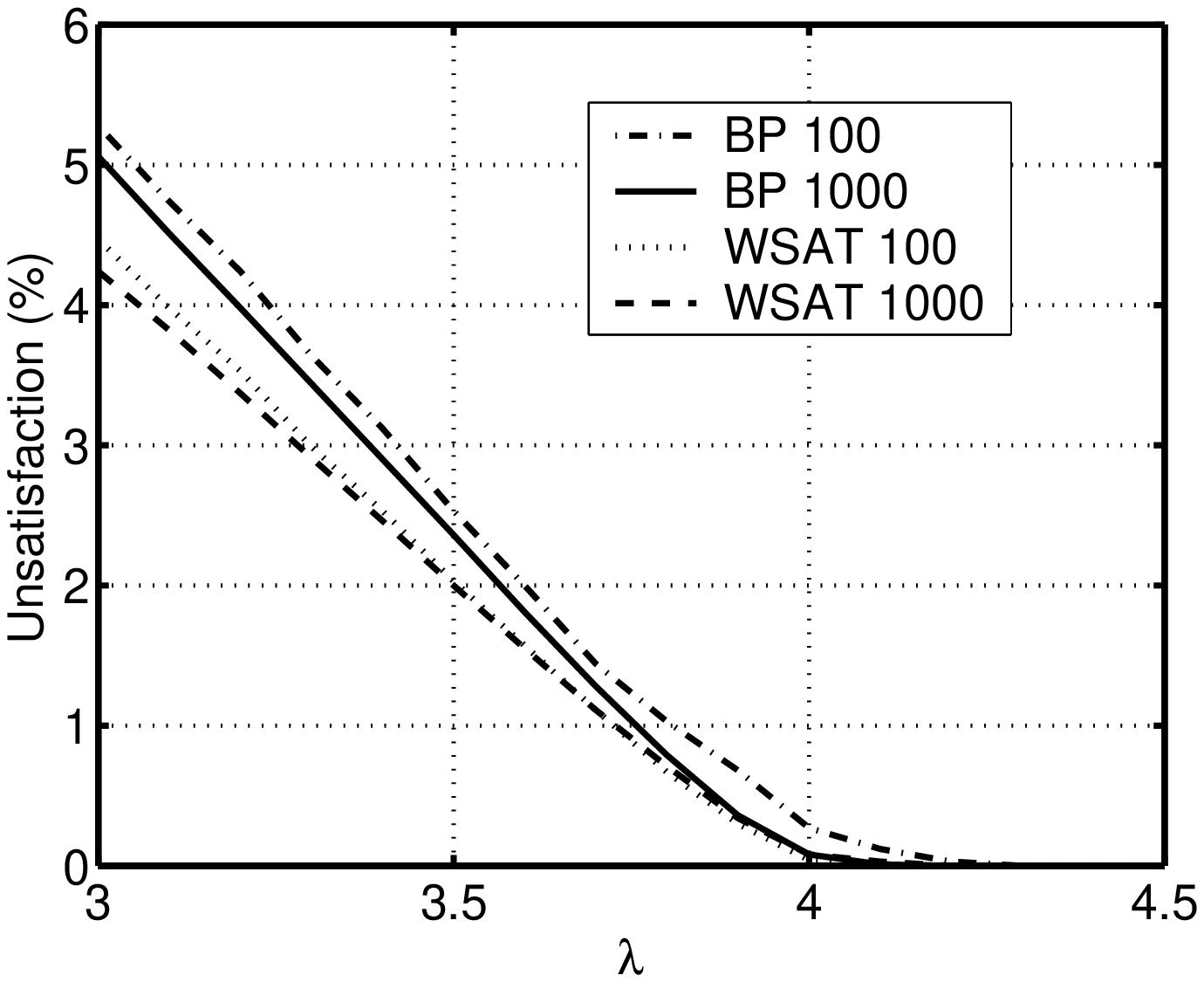}
\includegraphics[width=0.45\linewidth]{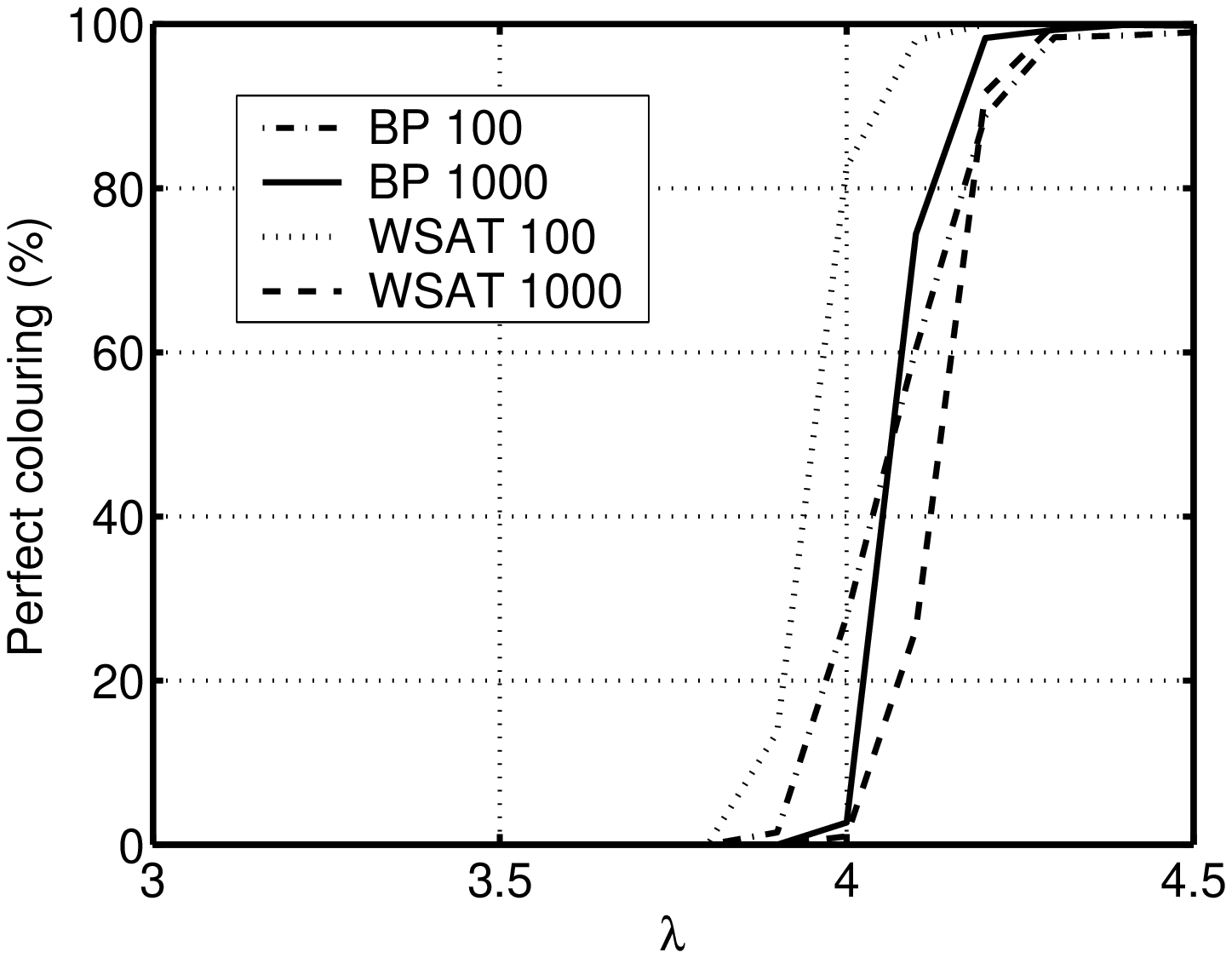}
\end{picture}
\caption{Comparison of belief propagation and Walksat algorithms for linear
graphs. (a) Unsatisfaction measure.(b) Percentage of perfect colouring. Note
that the percentage of perfect colouring solutions is 0 for Walksat on graphs
with 1000 vertices.}
\label{comp}
\end{figure}

\begin{figure}
\centering
\begin{picture}(400,180)
\put(10,160){$(a)$}
\put(220,160){$(b)$}
\includegraphics[width=0.43\linewidth]{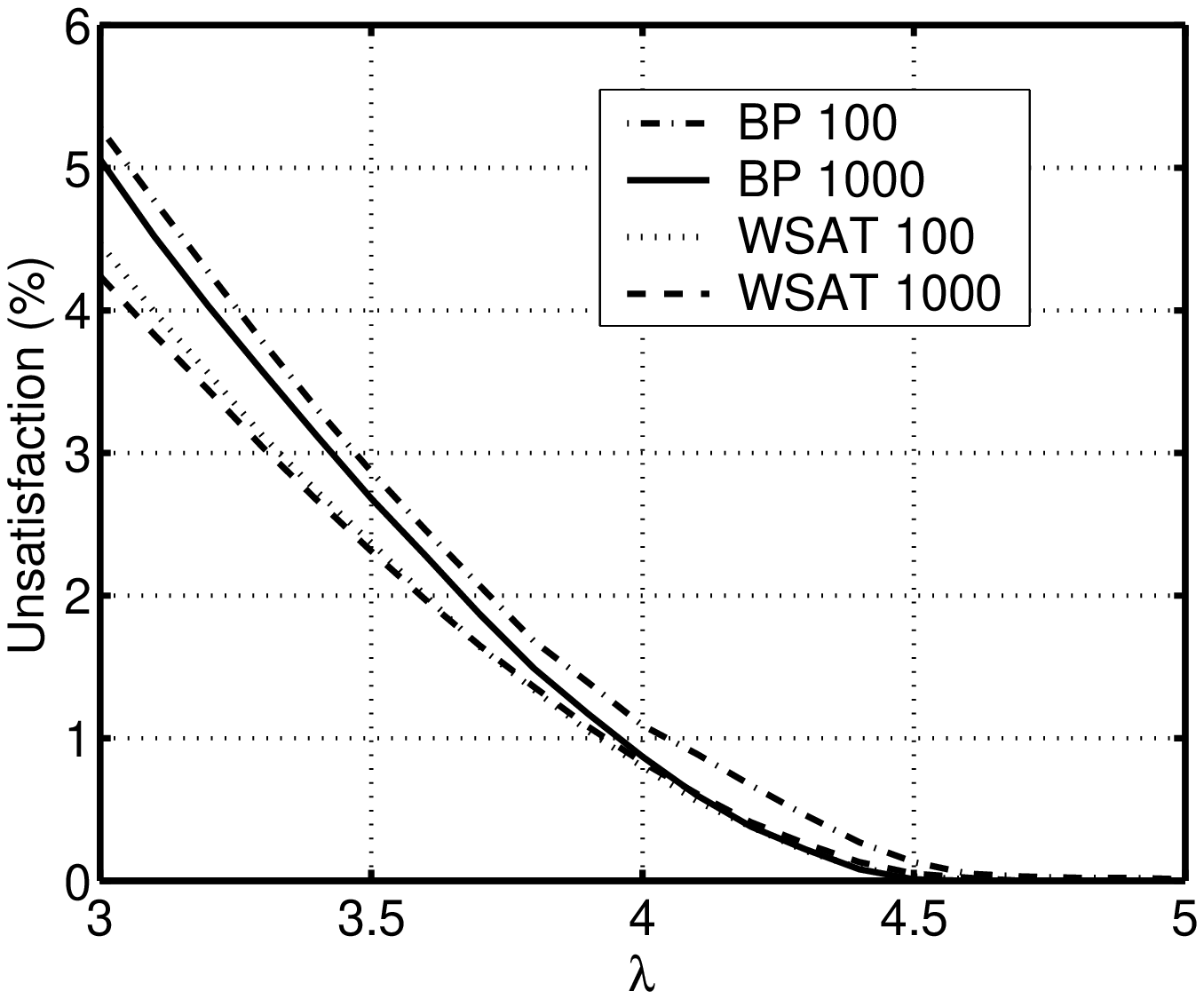}
\includegraphics[width=0.45\linewidth]{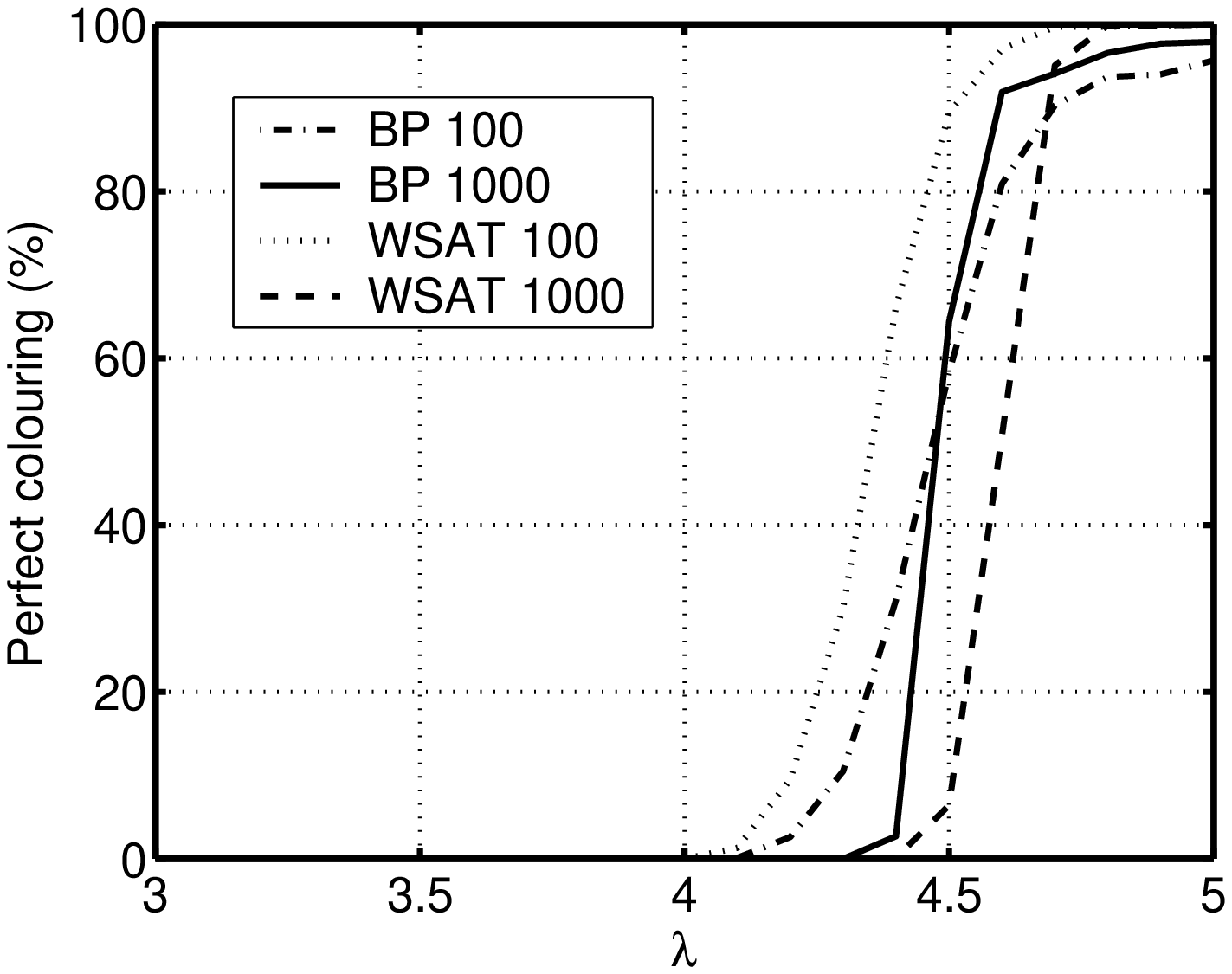}
\end{picture}
\caption{Comparison of belief propagation and Walksat algorithms for
Poissonian graphs. (a)~Unsatisfaction measure. (b)~Percentage of perfect
colouring.}
\label{comp_pois}
\end{figure}

Figure~\ref{comp} and~\ref{comp_pois} show first that the BP algorithm is
overall outperformed by the Walksat algorithm. We believe that the presence of
the small loops discussed in Sec.~\ref{sec:graph} is the main cause that
degrades the performance of the BP algorithm. Generalised belief propagation
method~\cite{gbp} is currently being investigated to obtain improved
performance. However, it also shows that the approximative BP algorithm works
surprisingly well considering the crude approximation made in (\ref{PZA}).

Then, it can be noticed that if Walksat clearly outperforms BP for 100 nodes
systems both in terms of unsatisfaction and of perfect colouring, this is not
the case when 1000 nodes systems are considered. Indeed, the results obtained
by BP seem better in terms of perfect colouring in that case for both linear and
Poissonian graphs. Obviously, if $nbit$ were to be increased significantly,
the Walksat algorithm would naturally outperform BP as for 100 nodes systems.
However, for an incresing system size and given computing resources BP is 
likely to outperform Walksat.

\section{Discussion}
\label{sec:conclusions}
We have studied a variation of graph colouring on random graphs of finite
average connectivity, aimed at maximising the number of colours accessible by a
vertex within one edge distance. The methods has significant practical
relevance, especially in the area of distributed storage that can be mapped onto
this problem. Two efficient algorithms, belief propagation and Walksat have been
adapted to carry out the task.

We have presented experimental results based on two types of random graphs for
different system sizes and identified the transition point,
in terms of the connectivity, for both algorithms to find a perfect solution.
For $q=4$ colours, we have found that the critical connectivity is around
$\lambda=4$ for linear graphs and around $\lambda=4.4$ for Poissonian graphs. 
In principle, the methods presented here can be used for random graphs
of any connectivity profile and any number of colours.

We have found that both algorithms give qualitatively very similar results, and
that the overall computer time needed to generate all the data presented here
was roughly the same for both algorithms. The relative efficiency of both
algorithms, in terms of the quality of obtained solutions and computing time, 
does however depend on the combination of parameters ($\lambda,~q,~n$) and graph 
characteristics. A more detailed analysis of this will be the subject of a 
separate study, as will be the thermodynamic phase diagram for this model.
Further research will focus on improving the message passing approach by using
the exact cluster expansion in the large system limit (i.e. focusing on stars 
and edges as our fundamental clusters instead of stars and nodes), combined 
with generalised BP, which will also provide us with a phase diagram for the 
model. It is assumed that this approach will remove the influence of short 
loops and therefore improve the performance of the algorithm, especially at low 
connectivity values.

If replica symmetry turns out to be broken, Survey Propagation~\cite{SP} like 
algorithm may further improve the results. Furthermore, we will consider to 
increase the distance to which a vertex can extend its search to retrieve the 
missing colours (e.g. second nearest neighbours), which will obviously change 
the basic clusters needed in the cluster expansion.

\ack Support from EU-FP6 IP EVERGROW is gratefully acknowledged.

\section*{Reference}

\end{document}